\documentclass[conference]{IEEEtran}
\IEEEoverridecommandlockouts
\usepackage{cite}
\usepackage{amsmath,amssymb,amsfonts}
\usepackage{algorithmic}
\usepackage{graphicx}
\usepackage{textcomp}
\usepackage{xcolor}
\usepackage{listings}
\usepackage{nopageno}
\usepackage{fancyhdr}
\def\BibTeX{{\rm B\kern-.05em{\sc i\kern-.025em b}\kern-.08em
    T\kern-.1667em\lower.7ex\hbox{E}\kern-.125emX}}

\begin{document}

\thispagestyle{empty}

\begin{huge}
IEEE Copyright Notice
\end{huge}

\vspace{5mm} 

\begin{large}
Copyright (c) 2020 IEEE
\end{large}

\vspace{5mm} 

\begin{large}
Personal use of this material is permitted. Permission from IEEE must be obtained for all other uses, in any current or future media, including reprinting/republishing this material for advertising or promotional purposes, creating new collective works, for resale or redistribution to servers or lists, or reuse of any copyrighted component of this work in other works.
\end{large}

\vspace{5mm} 

\begin{large}
\textbf{Accepted to be published in:} 6th Annual Conf. on Computational Science \& Computational Intelligence (CSCI'19); Dec 05-07, 2019; Las Vegas, Nevada, USA;

https://american-cse.org/csci2019/\#!/home

\end{large}

\vspace{5mm} 

\begin{large}
DOI 10.1109/CSCI49370.2019.00033
\end{large}

\fancyhead[L,C]{}
\fancyhead[R]{2019 International Conference on Computational Science and Computational Intelligence (CSCI\textbf{)}}
\renewcommand{\headrulewidth}{0.4pt}

\title{Plunge into the Underworld: A Survey on Emergence of Darknet}

\makeatletter
\newcommand{\linebreakand}{%
  \end{@IEEEauthorhalign}
  \hfill\mbox{}\par
  \mbox{}\hfill\begin{@IEEEauthorhalign}
}
\makeatother
\author{
\IEEEauthorblockN{Victor Adewopo}
\IEEEauthorblockA{\textit{School of Information Technology} \\
\textit{University of Cincinnati}\\
Cincinnati, Ohio, USA \\
adewopva@mail.uc.edu}
\and
\IEEEauthorblockN{Bilal Gonen}
\IEEEauthorblockA{\textit{School of Information Technology} \\
\textit{University of Cincinnati}\\
Cincinnati, Ohio, USA \\
bilal.gonen@uc.edu}
\linebreakand 
\IEEEauthorblockN{Said Varlioglu}
\IEEEauthorblockA{\textit{School of Information Technology} \\
\textit{University of Cincinnati}\\
Cincinnati, Ohio, USA \\
varlioms@mail.uc.edu}
\and
\IEEEauthorblockN{Murat Ozer}
\IEEEauthorblockA{\textit{School of Information Technology} \\
\textit{University of Cincinnati}\\
Cincinnati, Ohio, USA \\
ozermm@ucmail.uc.edu}
}

\maketitle

\begin{abstract}

The availability of sophisticated technologies and methods of perpetrating criminogenic activities in the cyberspace is a pertinent societal problem. Darknet is an encrypted network technology that uses the internet infrastructure and can only be accessed using special network configuration and software tools to access its contents which are not indexed by search engines.
Over the years darknets traditionally are used for criminogenic activities and famously acclaimed to promote cybercrime, procurements of illegal drugs, arms deals, and cryptocurrency markets. In countries with oppressive regimes, censorship of digital communications, and strict policies prompted journalists and freedom fighters to seek freedom using darknet technologies anonymously while others simply exploit it for illegal activities. Recently, MIT's Lincoln Laboratory of Artificial Intelligence augmented a tool that can be used to expose illegal activities behind the darknet.
We studied relevant literature reviews to help researchers to better understand the darknet technologies, identify future areas of research on the darknet and ultimately to optimize how data-driven insights can be utilized to support governmental agencies in unraveling the depths of darknet technologies. This paper focuses on the use of internet for crimes, deanonymization of TOR-services, darknet a new digital street for illicit drugs, research questions and hypothesis to guide researchers in further studies. Finally, in this study, we propose a model to examine and investigate anonymous online illicit markets.\footnote{This paper was presented in 6th Annual Conference on Computational Science \& Computational Intelligence (CSCI'19); Dec 05-07, 2019; Las Vegas, Nevada, USA; https://americancse.org/events/csci2019, 

(c)2019 IEEE

DOI 10.1109/CSCI49370.2019.00032}

\end{abstract}


\section{Introduction}
Darknet is referred to as the encrypted part of internet characterized with illegal and criminal activities. Darknet gained recognition in 1971, when two students of Massachusetts Institute of Technology and Stanford Universities traded illegal drugs (i.e., marijuana) using the Advanced Research Projects Agency Network (ARPANET) \cite{buxton2015rise}. There are often misconceptions with the terms ``Surface web'', ``Deep web'', and ``Dark web'' which are relatively interconnected but does not mean exactly the same thing. 

The surface web are web pages that are unencrypted and can be accessed using traditional search engines (e.g. Google, Bing, Yahoo). The surface web consists of billions static webpages and it occupies only 10\% of the internet space \cite{wikipedia}.

The dark web is a layer within the deep web and is not accessible using a standard browser. A number of techniques and tools have been designed to understand and crawl the deep web. A recent report indicated that the low harvest rate of deep web, about 647,000 distinct web forms, were found by sampling 25 million pages using Google index which constitutes only 2.5\% \cite{Zhao2016}\cite{dragut2012deep}. Deep web are contents available on the web that cannot be indexed or accessed by using search engines but can be accessed using a specific URL address. The deep web contains about 96\% of contents available on the internet \cite{Zhao2016}. 
\begin{figure}
  \includegraphics[width=7cm, height=3cm, ]{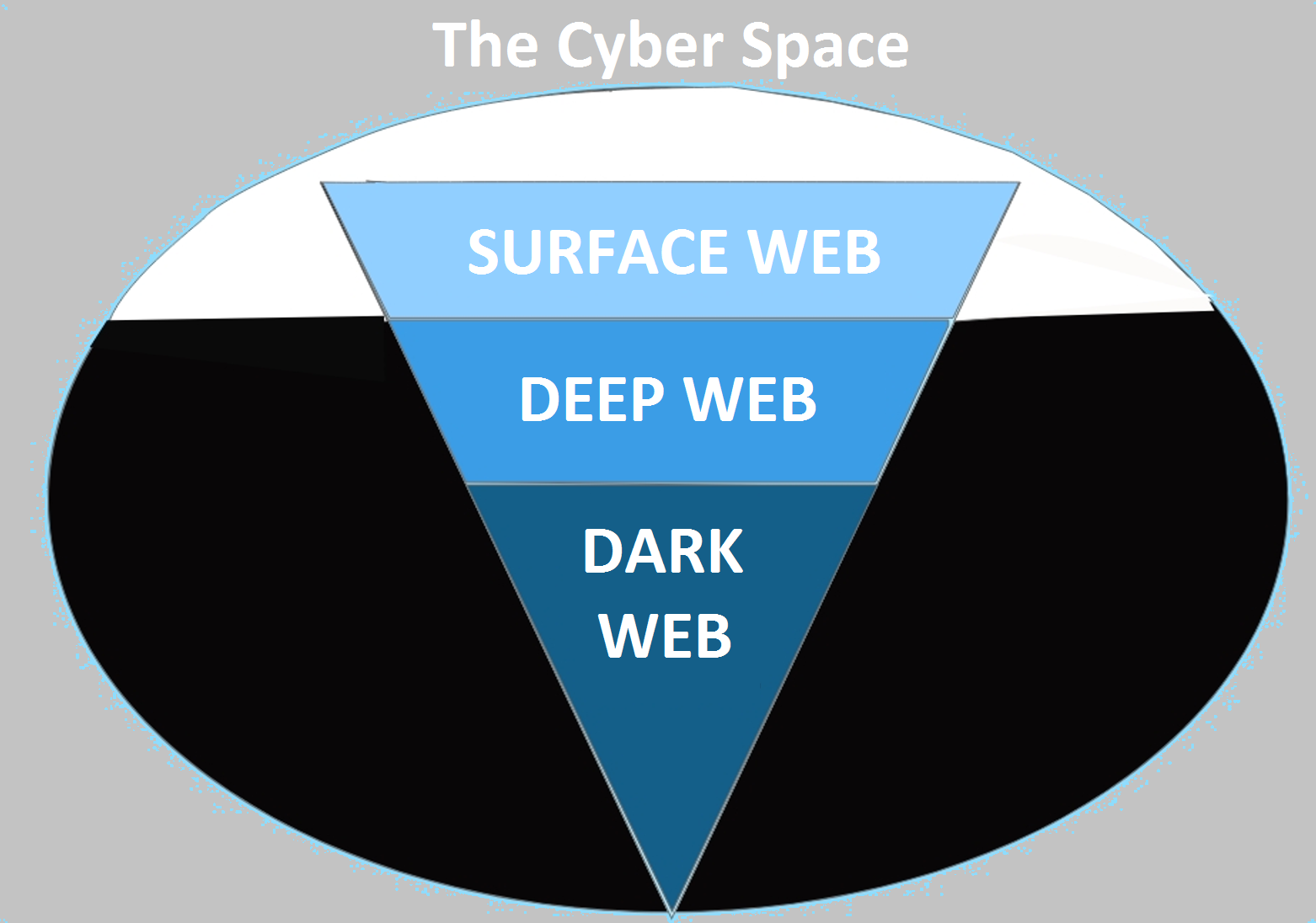}
  \centering
  \caption{Surface Web, Deep Web, And Dark Web Explained \cite{promptcloud}}
  \label{fig:f1}
\end{figure}
Darknet is an encrypted network technology that uses the internet infrastructure and can only be accessed using special network configuration and software tools to access its contents \cite{Mansfield-Devine2009}. The darknet uses peer-to-peer connection and privacy networks such as I2p, Tor \& Freenet to ensure the anonymity of users \cite{Bazli2017}. Traditionally, darknet is used for criminal activities, cybercrime, procurement of illegal drugs, child pornography, cryptocurrency markets, and hacking. The use of cryptocurrencies aided the sales of illegal and counterfeit products.Cryptocurrency (i.e., Bitcoins) is adopted as a method of payment in darknet market because of its’ anonymous characteristics which make it more difficult to link the users with any specific bank account details. A Cybersecurity firm ``UpGuard'' reported 419 million Facebook user's data breach in April 2019. Data exposed publicly on the Amazon server includes; passwords, user IDs, and check-ins \cite{Komando}. Similarly, in September 2018, over 50 million Facebook users' breached data was auctioned at a bitcoin value of 3\$ per each user's data on the darknet \cite{independent.co.uk}.

\section{Criminal Activities On Darknet}
\subsection{Use of Darknet for Internet Crimes}
The danger inherent in Tor services is significantly high compared to the perceived benefit. Oppressive regimes increasingly wants to monitor all digital communication, however lack of national laws and international cooperation by agencies is a big challenge in combating cyber-crimes on darknet websites \cite{Everett2015}. Tor is a software that ensures user's anonymity using three (3) relay nodes. Entry relay serves as a point of entry to the Tor network. The middle relay transports traffic from an entry relay to the exit relay. This is used to ensure anonymity and bridge gap between the entry and exit relay. The exit relay is the final relay that Tor traffic passes through before it reaches it's destination. The IP address of the exit relay is interpreted as the source of the traffic.
The Tor software emerged through the US-government funded The Onion Routing project (TOR) in 2001. Unethical contents are significantly high in darknet websites and users are inclined to engage more in unethical activities through support received in Tor service communities \cite{Guitton2013}. 

In October 2019, US department of justice charged two residents of Utah and 337 site users in a worldwide take down of largest darknet child pornography website called ``Welcome to Video'' \cite{q13fox}.
Cyberspace is a platform for us to share ideas and create improvements within our society, the increment in the utilization of cyberspace is faced with vulnerabilities. The increased rate on the use of cyberspace by fraudsters to generate malicious activities such as spam, identity theft DDoS, DRDoS requires a swift intervention. DDoS is the largest amount of cyber threat with over 48\% of cyber-attacks \cite{Fachkha2015}.

An alarming DDoS attack hit a peak of 400Gbps using a Network Time Protocol (NTP), CompTIA research revealed that a greater percentage of security breaches comes through inadvertent oversight of non-technical staffs. A novel approach is proposed to train and raise awareness of end-users on the overlooked basic security measures\cite{cloudflare} \cite{Furnell2009}.
 
Darknet can be used in monitoring cyber threats through deployment of technology visualization and image processing techniques. Darknet based monitoring system is designed to trap unused IP addresses with non-interactive hosts as a source of investigating and gathering intelligence \cite{Fachkha2016}
 
 \subsection{Deanonymization of Tor Services}
 Tor services are considered one of the most popular anonymizing services. The survey of Saleh et al.,\cite{Saleh2018} classified anonymity and security as the highest recurring keywords associated with TOR services in 26 years of research on TOR services. Identification of users' anonymity is a common phenomenon that has been studied by researchers. Many kinds of research have been carried out on deanonymization and development of new algorithms to crawl the Tor services, over 55\% of Tor researches focused on deanonymization and only 25\% of the studies focused on performance analysis and improvement. Studies show that differentiation of Tor services can ultimately be used to block TOR traffic\cite{winter2012china}
 
The Dark Web may perhaps be smaller compared to what we all project. The research of Owenson et al.\cite{Owenson2018} seeks to identify the actual size of the darknet services, the study illustrated that the dark web is not as huge as many researchers perceived. The use of web crawlers was implemented to retrieve the contents of long-lived and short-lived hidden services. It was revealed that more than half of the services discovered disappeared within 24 hours. Furthermore, HTTP is commonly used for longer-lived onions and zeroNet for the shorter-lived onions \cite{kadianakis2015extrapolating}.
 
Over the years, closure of a new darknet service leads to evolution of another in a new form \cite{soska2015measuring}. The darknet is a dynamic and agile environment for market activities, vendor activities declined after operation onymous which lead to the termination of 410 hidden services and arrest of site vendors, over time criminal activities progressed geometrically in a new form \cite{decary2017police}.
Lane et al. studies analyzed market activities to understand vulnerabilities to the interference of darknet markets using EAST (Event Analysis of systemic Teamwork) broken-links. This approach was proposed as a key component that law enforcement agencies can use to target the intentional disruption of services. \cite{Lane2019}
The use of IoT devices including wearables, smart devices, and smart homes has proven beneficial to human activities. Utilization of these devices is on the rise and is vulnerable to cyber-attack. A study on a famous malware of the IoT was conducted to propose new methods of detecting cyberattacks before they occur. Ozawa et al. study focused on TCP (Transfer Control Protocol) and SYN (Synchronization flag) packets to characterize scan attacks. A new approach was proposed using Association Rule Learning to scan destination ports \& TCP/IP header information for attacks on packets found on darknets. The approach was successful in detecting malware attacks before they occur \cite{Ozawa2019}.

The availability of sophisticated technologies for cybercrimes has risen to a high level of pertinent concern and societal problems. The anonymity of actors in cyberspace facilitates freedom of speech, privacy and empowers nefarious actors to perpetrate illegal activities \cite{Lane2019}.

Benjamin et al. developed a framework for identification of darknet data, method of data collection, evaluation and ethical concerns in researching darknets \cite{benjamin2019dice}.  Haughey et al. developed an algorithm capable of reducing the anonymity of Tor users. Adaptive Traffic Fingerprinting for Darknet Traffic Intelligence (ATABI) with HTML injection, Border Gate Protocol (BGP) routing attack, and a Detection technique was perpetrated against a simplified version of (MITM) Man in the Middle Attack which yielded a positive response, while low sensitivity was observed in non-targets without MITM \cite{Haughey2018}.

Fidalgo et al. developed an approach to support law enforcement agencies to discover evidence of criminal activities on darknet. The approach deployed the automatic classification of digital images uploaded on darknet websites using Bag of Vision Words (BOVW) to eliminate pixels of background and foreground images of interest. A filtering strategy, Semantic Attention Key-point Filtering (SAKF) is compared against a well-known convolutional neural network to detect images with illicit visual contents \cite{Fidalgo2019}.

Dalins et al. also developed a Tor-use Motivation Model (TMM) to explicitly distinguish sites contents from motivations in Tor websites. The TMM utilize bots to crawl web-pages of Tor services, which was classified into illegal activities, child pornography, and narcotics. This model was developed to help law enforcement in prioritizing and appraisal of evidence for further investigations. Majority of the contents on the pages revolve around criminal activities (Narcotics and illegal financial services) \cite{Dalins2018}.

Pang et al. proposed a method of identifying malicious activity through traffic flow of darknet in identifying groups of malicious events from historical darknet traffic. The behavioral grouping of malicious activities on darknet traffic is a useful tool in observing patterns of cyber threats \cite{Pang2017}.

Darknet is a secretive anonymous place, the use of the services for legitimate or illegal activities is a factor centered on individual actors. The Tor services concept of entering the network through a relay with details packaged by hidden services with public key sent to hash tables on exit, networks are sent to a different IP making it untraceable is filled with vulnerabilities, law enforcement agencies can run sponsored Tor relays to lure unsuspecting actors to utilize the services \cite{Bradbury2014}.

\section{Darknet, Digital Street for Illicit Drugs}
Crypto markets are online market place on Darknet websites which provides an anonymous platform for trading of illicit drugs and services. Over 75\% of darknet market activities are associated with illegal drug deals with specific goods being sought for in several originating and specific destination countries \cite{Broseus2017} \cite{Goosdeel2017}. Perfect markets do not exist in darknet cryptomarkets, communities of interest participate in negotiation and supply of drugs which have a significant impact on street sales of drugs. Stimson's \& Gerry studies indicated a positive relationship between the demand rates of drugs and cost in a non-participant observation and in-person interview of darknet users \cite{Stimson2017}.

The first illicit drug darknet market Silk Road with an approximate life span of 2 years; opened in January 2011 and was seized by the US Federal Bureau of Investigation (FBI) on October 2013. Shortly after, Silk Road 2.0 was launched to continue market activities. Several drug market activities have emerged on darknet websites and are functional till date, darknet marketplaces-maintained anonymity of users through the use of cryptocurrencies introduced in 2008 \cite{Dolliver2015}.

Espinosa et al. study accessed impact of e-reputation on online darknet markets. Vendors often build a good reputation online while vendors with bad reputation close down their account and reopen a new one with another identity which is synonymous to activities on clearnet markets. The E-reputation of sellers is used to signal the quality of products and build trust with sellers in darknet markets \cite{Espinosa2019}\cite{RachelBotsman2017}. Three performance metrics often used to identify the quality of drugs in darknets include; purity, chemical knowledge of components and embodied experience \cite{Bradbury2014}.

A theoretical figure of the configuration used to gather communications between actors in five darknet markets identified external threats relating to risks in darknet markets. Actors do not fully trust communication between each other as possibilities of law enforcement agencies could disguise to gather intelligence from unsuspecting users \cite{Kamphausen2019}.

Only a few studies explored the impact of illegal drug activities on healthcare. The upsurge of illegal drugs, child trafficking, and pedo-pornography have a sheer impact on environmental wellbeing. More than 11.4 million people misused opioids in 2018, and in 2017 over 70,000 deaths occurred due to drug overdose \cite{OftheAssistantSecretaryforHealth2019}. 300\% increments in deaths due to drug overdose was observed between 1997 and 2013 \cite{Wunsch2009}.
Some hackers have successfully been able to hack medical data which are sold on darknet markets or siphoned fake medical insurance claims. Vulnerabilities in healthcare data can be reduced through effective synergism of effort between engineers, medical professionals, and other professionals \cite{Masoni2016}.

The distribution of illegal drugs via darknets allows the marketers of drugs to locate potential buyers and send drugs via legitimate means such as mail posts without been traced. New accessibility to drugs on Tor services increased the rate of patients abusing drug use \cite{Pergolizzi2017}. Cunliffe et al. study collected data over a period of three years using data crypto software through web crawling in over thirty-one darknet markets addressed the purchase of Non-medical Prescription Psychiatric Drug Use (NMPDU). The occurrence of NMPDU drugs are significantly low on surveyed markets, however, the sales of sedatives and stimulants are on the far high side in US \cite{Cunliffe2019}.

Soska and Christin \cite{soska2015measuring} examined the online anonymous marketplace ecosystem for four years. According to their research, ``Silk Road'' was one of the most popular darknet websites in the world after 2011. Silk Road announced that it was an anonymous marketplace where people can find whatever they want, just like eBay. However, Silk Road provided stronger anonymity with Tor hidden services which use a decentralized network. Silk Road has a rating system for buyers and sellers. Also, it was widely used as smuggling including drug and weapon selling way, and it was trendy among criminals. Finally, its provider was arrested, and the website has been shut down by the government and police. Buyers and sellers moved to another darknet website, that is Silk Road 2.0. Besides, many darknet websites used the same marketplace method. In the evolving period, many deep web marketplaces have experienced police raids or court closure orders. They demonstrate how real-world criminals develop an ecosystem on the internet. Illegal drug selling or buying is 70\% of all sales. There is no physical transaction among sellers, buyers and marketplace owners. Police was not able to track the transactions. Silk Road used Bitcoin's multisig feature that ensures transaction validation. Furthermore, Silk Road provided a feedback system to measure quality control on the sales. Evolution and Agora marketplaces which are popular darknet platforms used same methods. The authors collected 35 marketplace data. They observed that marketplaces have pretty low reliability under 70\%. They found and analyzed 79,512 products, and about 5\% of them have very high prices. In Silk Road, the amount of Bitcoin transactions was \$300,000 per day. Thus, it could reach \$100 million US dollar within a year. There are a couple of related works. Finally, the paper demonstrated that Silk Road made \$650,000 daily transaction. The authors recommended that there is an urgent need to re-evaluate the anonymous online marketplaces. 

Broséus et al. \cite{Broseus2017} examined ``Evolution'' online anonymous marketplace. They collected 92,980 unique sale proposal with 4171 distinct vendors in 2014 and 2015. The study demonstrated that a vendor handle 22 products on average. 63\% of all sales were illicit drugs and paraphernalia in all the countries. 69\% of all vendors were selling those illicit drugs. On the other hand, about 1,000 out of 92,980 products were firearms. There were 93 shipping countries with 164 distinct destinations. 
26.1\% of all vendors were shipping their products to United States. 32\% of all illicit drugs were shipping to United States. In 2014, 53\% of all illicit drugs sales is Cannabis for United States. The top 5 shipping destination countries are United States, United Kingdom, Germany, Netherlands and Australia. The results suggested that some vendors had big number of products and dominate the darknet market. Also, this website provided ID theft including driving licenses, passports, illegitimate documents such as fake bank statements or university diplomas, fake credit card numbers, e-mail, social security number, date of birth, mother’s maiden name. 

\section{DISCUSSION AND FUTURE WORK}

With the comprehensive literature review's light, it is clear that researching darknet activities is vital for public safety and the future of our community. Notably, drug overdose deaths are related to the darknet activities. Our hypothesis is that street illicit drug dealers are becoming darknet online illicit drug vendors. Also, since easy and untraceable shipping and payment system that relies on cryptocurrency transactions (especially Bitcoin), drug addicts tend to buy illicit drugs from darknet. Thus, our primary goal fundamentally is to research darknet illegal market transactions to help in reducing drug overdose incidents. The secondary goal is to reveal darknet secrets. We are considering examining three darknet platforms to dive deeply into the darknet world. We will conduct not only research with descriptive statistics but also with a revealing examination of traceability of darknet illegal market transactions. Thus, we will follow up with the research questions below in the next study.

\begin{enumerate}
\item What are the values embedded in information extracted from darknet websites through research in supporting government agencies, and how can that be optimized?
\item How can information sharing across agencies be improved to support investigations and prosecution strategies, whether to build an instant case or connect other potentially related cases?
\item How can we track down the illicit drug trafficking payments on cryptocurrency activities?
\end{enumerate}

We will offer a model that covers a combined technique to de-anonymize darknet related bitcoin transactions. The raw model will be consisted of ten steps. 

\begin{enumerate}
\item Method-1: Test proposed algorithms by researchers to de-anonymize bitcoin transactions used in darknet activities
\item Method-2: Identify darknet activities related to Bitcoin addresses and track the incoming and outgoing traffic
\item Method-3: Analyze for abnormalities in Bitcoin prices in order to detect likely darknet transactions 
\item Method-4: Automatically check real-time transactions to capture catched Bitcoin addresses used in previous darknet activities
\item Server-1: Monitor Bitcoin Ledger
\item Server-2: Blacklisted Bitcoin data analysis
\item Action-1: Open Source analysis through automated scans to find pseudonyms of darknet activities related to Bitcoin addresses
\item Action-2: Analytical analysis (link analysis) that constantly augments de-anonymization by associating previously unlinked Bitcoin addresses
\item Action-3: Manual data entry from different users/victims to increase de-anonymization process
\item Action-4: Constant feedback from the field to evolve the accuracy of the system
\end{enumerate}

\bibliographystyle{IEEEtran}
\bibliography{references}

\end{document}